\documentclass[prl,twocolumn,superscriptaddress,showpacs,amsmath,amssymb]{revtex4-1}

\newcommand{\dd}[1]{\mathrm{d}#1\,}
\newcommand{\DD}[1]{\mathrm{D}[#1]\,}
\newcommand{\avg}[1]{\langle{#1}\rangle}

\DeclareMathOperator{\Tr}{Tr}
\DeclareMathOperator{\tr}{tr}

\newcommand{\sgn}{\mathop{\mathrm{sgn}}}

\newcommand{\bra}[1]{\langle{#1}\rvert}
\newcommand{\ket}[1]{\lvert{#1}\rangle}

\renewcommand{\vec}[1]{\bm{#1}}

\newcommand{\arxiv}[2]{#1}

\usepackage{graphicx}
\usepackage{latexsym}
\usepackage{amsmath}
\usepackage{amssymb}
\usepackage{amsfonts}
\usepackage{color}
\usepackage{units}
\usepackage{bm}
\usepackage{verbatim}
\usepackage{hyperref}

\begin{document}

\title{Spin pumping and torque statistics in the quantum noise limit}

\author{P.~Virtanen}
\affiliation{NEST, Istituto Nanoscienze-CNR and Scuola Normale Superiore, I-56127 Pisa, Italy}
\affiliation{University of Jyvaskyla, Department of Physics and Nanoscience Center, P.O. Box 35, 40014 University of Jyv\"askyl\"a, FINLAND}

\author{T.T.~Heikkil\"a}
\affiliation{University of Jyvaskyla, Department of Physics and Nanoscience Center, P.O. Box 35, 40014 University of Jyv\"askyl\"a, FINLAND}

\begin{abstract}
We analyze the statistics of charge and energy currents
and spin torque in a metallic nanomagnet coupled to a large magnetic
metal via a tunnel contact. We derive a Keldysh
action for the tunnel barrier, describing the stochastic currents in
the presence of a magnetization precessing with the rate
$\Omega$. In contrast to some earlier approaches, we include the
geometric phases that affect the counting statistics. We
illustrate the use of the action by deriving spintronic fluctuation
relations, the quantum limit of pumped current noise, and consider the
fluctuations in two specific cases: the situation with a stable
precession of magnetization driven by spin transfer torque, and the
torque-induced switching between the minima of a magnetic
anisotropy. The quantum corrections are relevant when
the precession rate exceeds the temperature $T$, i.e., for $\hbar
\Omega \gtrsim k_B T$. 
\end{abstract}

\maketitle

Spin transfer torque, angular momentum contributed by electrons
entering a magnet, can be used to control magnetization dynamics via
electrical means, as demonstrated in many experiments.
\cite{slonczewski1996-cde,tserkovnyak2005-nmd,ralph2008-stt} Often the effect
can be described by considering the ensemble average magnetization
dynamics, or taking only thermal noise into account. \cite{brown1963-tfs}
The spin transfer torque is in general also a stochastic process, but at bias
voltages large enough to drive the magnetization, it is not
necessarily Gaussian nor thermal, \cite{foros2005-mnm} especially at
cryogenic temperatures.  The statistical distribution of electron
transfer and the associated torque in magnetic tunnel junctions can be described by
counting statistics, \cite{levitov1993-cdi} via a joint probability
distribution of charge, energy, and spin transferred into the magnet
during time $t_0$, $P_{t_0}(\delta{}n,\delta{}E,\delta\vec{}s)$. The
distribution is conditional on the magnetization dynamics during time
$t_0$, which necessitates consideration of back-action effects.

\begin{figure}
  \includegraphics{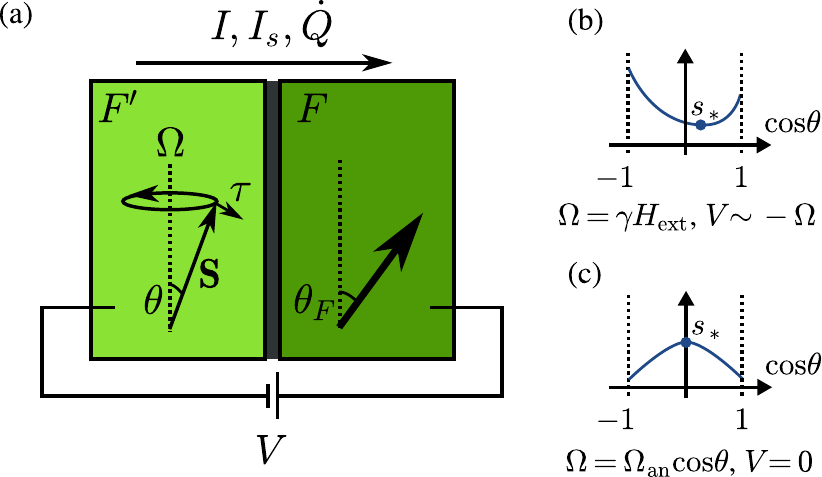}
  \caption{\label{fig:setup}
    (a)
    Tunnel junction between magnetic materials
    with free ($F'$) and fixed ($F$) magnetizations.
    The total spin $\vec{S}={\cal V}\vec{M}/\gamma$ in $F'$ 
    precesses at angular frequency $\Omega$ around the $z$-axis.
    As described by Eq.~\eqref{eq:genfunc},
    the motion pumps charge, spin and heat currents through the junction,
    and the back-action spin transfer torque $\vec{\tau}$
    drives a change in the tilt angle $\theta$.
    (b--c)
    Schematic of effective magnetic potential energy, in the presence
    of an external field $H_{\rm ext}$ and large spin transfer torque,
    or, in the presence of a magnetic anisotropy $\Omega_{\rm an}=\gamma{}H_{\rm an}$.
  }
\end{figure}

Here we construct a theory describing the probability distribution for
electron transfer via a Keldysh action
(Eq.~\eqref{eq:tunnel-action}) describing a metallic magnet with
magnetization $\vec{M}$, coupled to a fermionic reservoir (another
ferromagnetic metal), illustrated in Fig.~\ref{fig:setup}. In the presence of a
bias voltage in the reservoir, this coupling may lead to a stochastic
spin transfer torque affecting the magnetization dynamics. Unlike some
of the earlier discussions of counting and spin torque statistics
\cite{utsumi2015-fts,chudnovskiy2008-sts,tang2014-fcs}, we follow
the approach of Ref.~\cite{shnirman2015-gqn} and retain geometric phase factors in the
derivation of the generating function. This becomes relevant in the
quantum limit $\hbar\Omega > k_B T$ where the precession rate $\Omega$ is
large compared to the temperature $T$.

To study the implications, we suggest two specific settings
(Fig.~\ref{fig:setup}b,c), characterized by opposite regimes of the
external field ${\mathbf H}_{\rm ext}$ and anisotropy field ${\mathbf
  H}_{\rm an}$. When $H_{\rm ext} \gg H_{\rm an}$, a suitably chosen
voltage drives the magnet into a stationary precession with rate
$\Omega$ around the direction of ${\mathbf H}_{\rm ext}$.
\cite{slonczewski1996-cde,berger1996-esw,kiselev2003-mon,rippard2004-dci}
This precession pumps charge \cite{tserkovnyak2002-egd} and heat into the reservoir, along with
the direct charge and heat currents due to the applied voltage.  The
noise of these currents depends on the intrinsic noise of the pumped
current and, at low frequencies, also on the fluctuations of the
magnetization, driven by the spin torque noise. The opposite limit
$H_{\rm an} \gg H_{\rm ext}$ is the one relevant for memory
applications, as the spin transfer torque can be used to switch
between the two stable magnetization directions
\cite{koch2004-trr,yakata2009-tss}. Our approach allows finding the
switching rate at any temperature and voltage, also for
$k_BT\ll\hbar\Omega$.

Besides the average currents and noise, the Keldysh action allows
us to calculate the full probability distribution $P_{t_0}(\delta
n,\delta E,\delta s)$ of transmitted
charge $\delta n$, energy $\delta E$, or change $\delta M_z={\cal
  S}\gamma \delta s/{\cal V}$ of the $z$-component of magnetization in
a nanomagnet with volume ${\cal   V}$ and spin ${\cal S} \gg 1$, within a long measurement time 
$t_0$. Here $\gamma$ is the gyromagnetic ratio. The precise
distribution depends on the exact driving conditions and the
parameters of the setup. However, symmetries
constrain the probability distribution, leading to a spintronic
fluctuation relation (here and below, $k_B=\hbar=e=1$)
\begin{equation}
\label{eq:fluctuationrelation}
\begin{split}
&P_{t_0}(\delta n,\delta E,\delta s)= \\& e^{V \delta n/T_F}
e^{\delta E (T_F^{-1}-T_{F'}^{-1})} e^{-\Omega\delta s/T_F} P_{\rm t_0}'(-\delta n,-\delta E,\delta s),
\end{split}
\end{equation}
where $P_{\rm t_0}'$ corresponds to the case with reversed
magnetizations. As in
fluctuation relations presented earlier \cite{crooks1999-epf,crooks2000-pea,seifert2012-stf,esposito2009-nff,tobiska05,utsumi2015-fts},
this allows for a direct derivation of Onsager symmetries,
thermodynamical constraints, and fluctuation-dissipation relations,
valid for the coupled charge-spin-energy dynamics \arxiv{(see Appendix)}{\cite{supp}}.

{\em Generating function.}
Consider a magnetic tunnel junction depicted in Fig.~\ref{fig:setup}.
The spin transfer torque due to
tunnelling, and the corresponding counting statistics can be described
by a Keldysh action obtained by integrating out conduction electrons
in $F$ and $F'$.
\cite{chudnovskiy2008-sts,shnirman2015-gqn}
We apply the approach of
Ref.~\onlinecite{shnirman2015-gqn} to the characteristic function
$Z(\chi,\xi)=\avg{e^{i[N_F(t_0)\chi_0+(H_F(t_0)-\mu)\xi_0]}e^{-i[N_F(0)\chi_0+(H_F(0)-\mu)\xi_0]}}$
describing the change in particle number $N$ and internal energy $H_F$ in
the ferromagnetic lead $F$.
\cite{esposito2009-nff,campisi2011-cqf} In the long-time limit,
$t_0\gg{}1/T,1/V$, this results to the action $S=S_0+S_T$, where
$S_0=\mathcal{S}\int_{-\infty}^\infty\dd{t}[-2\dot{\psi}^q-\sum_\pm(\pm\dot{\phi}^{cl}+\dot{\phi}^q)\cos(\theta^{cl}\pm\theta^q)]$
is the Berry phase for total spin
${\cal{}S}={\cal V}|\vec{M}|/\gamma$. Moreover,  the tunneling action is
\begin{align}
  \label{eq:tunnel-action}
  S_T
  &=
  i|W|^2
  \int_{-\infty}^\infty\dd{t}\dd{t'}\frac{\dd{\epsilon}}{2\pi}
  \Tr \check{P}(t) \check{G}_{F'}(t-t')\check{P}(t')^\dagger \check{G}_F(\epsilon)
  \,,
\end{align}
where $\check{P}(t) = e^{i(\epsilon-V)t}e^{i[\chi(t) +
    (\epsilon-\mu)\xi(t)]\check{\gamma}_x/2}\check{R}(t)$ contains the
bias voltage $V$, and the charge and energy counting fields $\chi(t)$
and $\xi(t)$. The
rotation matrix
$\check{R}(t)=e^{-i\check{\phi}(t)\sigma_z/2}e^{-i\check{\theta}(t)\sigma_y/2}e^{-i\check{\psi}(t)\sigma_z/2}$
describes the direction of the magnetization
$\vec{S}=(\cos\phi\sin\theta,\sin\phi\sin\theta,\cos\theta){\cal S}$
in terms of Euler angles $\theta$ and $\phi$.
Keldysh fields are in the basis \cite{kamenev2010-kta}
$\check{\phi}=\phi^{cl}+\phi^q\check{\gamma}_x$, where
$\check{\gamma}_x$ is a Pauli matrix.  Below, we fix the gauge 
\cite{shnirman2015-gqn} so that $\psi^q=-\phi^q\cos\theta^{cl}$,
$\dot{\psi}^{cl}=-\dot{\phi}^{cl}\cos\theta^{cl}$. We assume
a spin and momentum independent tunneling matrix element $W$.  The
conduction electrons are described by
Keldysh--Green functions $\check{G}$, with the exchange field of $F'$
always parallel to $\hat{z}$ in the rotating frame,
$\check{G}_{F'}^R(\epsilon,\vec{k})=[\epsilon-\xi_{\vec{k}}+h_{F'}\sigma_z]^{-1}$.

Consider now the situation depicted in Fig.~\ref{fig:setup}a, where
$\vec{S}$ precesses around $\hat{z}$ due to an external magnetic
field and/or magnetic anisotropy contributing potential energy $E_M$. The corresponding action
is
$S_{\rm{}ext}=\int\dd{t}\sum_{\pm}\pm{}E_M[\vec{S}_\pm]=2\int\dd{t}\vec{S}^q\cdot\hat{z}\Omega$,
with $\Omega=\Omega_{\rm ext} + \Omega_{\rm an}[\cos\theta]^{cl}$.
Separating out the fast motion $\phi^{cl}(t)=\Omega
t+\tilde{\phi}^{cl}(t)$, the dynamics of $\theta$, $\tilde{\phi}$ are
driven only by the spin transfer torque.  We assume this dynamics is slow, and
evaluate Eq.~\eqref{eq:tunnel-action} under a time scale separation
$\sim{}t_0^{-1},|W|^2/{\cal S}\ll{}T,\Omega$:
\footnote{
  Here and below, we choose $\psi(t)$ as in~\onlinecite{shnirman2015-gqn},
  and only consider $\psi^q(\pm \infty)=0$.
}
\begin{align}
  \label{eq:genfunc}
  &S_T
  \simeq
  -i
  \int_{-\infty}^\infty
  \dd{t}
  \dd{\epsilon}
  \sum_{\sigma\sigma'\alpha=\pm}
  \Gamma_{\sigma\sigma'\alpha}(\epsilon) (e^{i\alpha\eta_{\sigma\sigma'}(\epsilon)} - 1)
  \,.
\end{align}
Here,
$\eta_{\sigma\sigma'}(\epsilon)=\chi(t)+(\epsilon-\mu+V+\Omega_{\sigma\sigma'})\xi(t)-2\frac{\Omega_{\sigma\sigma'}}{\Omega}\phi^q(t)$ and $\Omega_{\sigma\sigma'} = [\sigma\Omega-\sigma'\Omega\cos\theta^{cl}(t)]/2$.
The transition rates per energy are
\begin{align}
  \Gamma_{\sigma\sigma'\alpha}(\epsilon)
  &=
  \bar{G}_{\sigma\sigma'}
  \frac{1 + \sigma\sigma'\cos\theta^{cl}(t)}{2}
  \Lambda_{\alpha}(\epsilon,V+\Omega_{\sigma\sigma'})
  \,,
  \\
  \Lambda_{\alpha}(\epsilon,V)
  &=
  \begin{cases}
    f_{F'}(\epsilon)[1 - f_{F}(\epsilon+V)]\,, & \alpha = + \,,
    \\
    f_{F}(\epsilon+V)[1 - f_{F'}(\epsilon)]\,, & \alpha = - \,.
  \end{cases}
\end{align}
Here, $f_{F/F'}(\epsilon)=1/[e^{(\epsilon-\mu)/T_{F/F'}}+1]$ are Fermi
distribution functions, and the time-averaged conductance is
$\bar{G}_{\sigma\sigma'}=G_0\frac{1+\sigma{}P_{Fz}}{2}\frac{1+\sigma'P_{F'}}{2}$
where
$G_{0}=2\pi|W|^2(\nu_{F\uparrow}+\nu_{F\downarrow})(\nu_{F'\uparrow}+\nu_{F'\downarrow})$,
the polarizations are defined as
$P=(\nu_{\uparrow}-\nu_{\downarrow})/(\nu_{\uparrow}+\nu_{\downarrow})$,
and $P_{Fz}=P_F\cos\theta_F$ is the polarization of the fixed magnet
projected onto the precession axis.  The densities of states
$\nu_{\uparrow/\downarrow}$ of majority/minority spins are given at
the Fermi level.
The resulting $S_T$ is independent of $\tilde{\phi}^{cl}$, i.e. its
dynamics decouples, which constrains $\theta^q=0$ \arxiv{(see Appendix)}{\cite{supp}}.

The result describes Poissonian transport events, each associated with
a back-action on $\theta$ due to the spin transfer torque, as
described by the dependence on $\phi^q$.  The rates are proportional
to the averaged densities of states and squared spin overlaps
$|\langle{\sigma}|\sigma'\rangle|^2=[1+\sigma\sigma'\cos\theta]/2$, in
the frame rotating with the magnetic precession. The transferred
energy $V+\Omega_{\sigma\sigma'}$ consists of the voltage bias
and the difference $\pm\Omega/2 -(\pm\Omega\cos\theta)/2$ of
energy shifts on the right and left sides of the junction in the
rotating frame \cite{tserkovnyak2005-nmd,tserkovnyak2008-tbe}. The
relation of this additional dependence on $\theta$  to geometric phases is
discussed in Ref.~\onlinecite{shnirman2015-gqn}.  It also separates
Eq.~\eqref{eq:genfunc} from the result of
Ref.~\onlinecite{utsumi2015-fts} for tunneling through a ferromagnetic
insulator barrier, where such angular dependencies are not included.

Equation \eqref{eq:genfunc} is a main result of this work, as
the knowledge of $S_T$ allows access to the statistics of
charge, energy and spin transfer in the generic case depicted in
Fig.~\ref{fig:setup}a. Below, we describe some applications. First, we can identify the
following spintronic fluctuation relation \arxiv{(see Appendix)}{\cite{supp}}
\begin{align}
  S_T(\chi,\xi,\phi^q)
  =
  S_T'(
  -\chi+\frac{iV}{T_F},
  -\xi +\frac{i}{T_F}-\frac{i}{T_{F'}},
  \phi^q+\frac{i\Omega}{2T_F})
  \,,
\end{align}
where the prime denotes inverting the magnetizations and the sign of
the precession. Identifying the conjugate fields of $\chi,\xi$, and $\phi^q$ to the
number of charges $\delta n$, change of energy $\delta E$ and transfer
of spin angular momentum $\delta s$, this relation is equivalent with
Eq.~\eqref{eq:fluctuationrelation}. This relation also implies the
Onsager relation $dI/d\Omega=(\sin^2\theta)d\tau/dV$
relating the pumped current to the torque $\tau\sin^2 \theta \equiv
dS_T/d(2\phi^q)$ acting on the angle $\theta$
\cite{brataas2011}. This and further details of the fluctuation
relation are discussed in \arxiv{the Appendix}{\cite{supp}}.

\begin{figure}
  \includegraphics{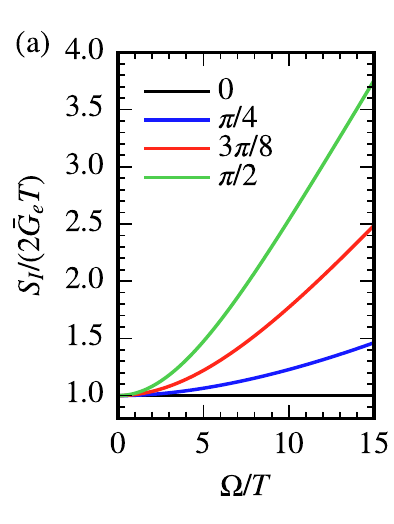}
  \includegraphics{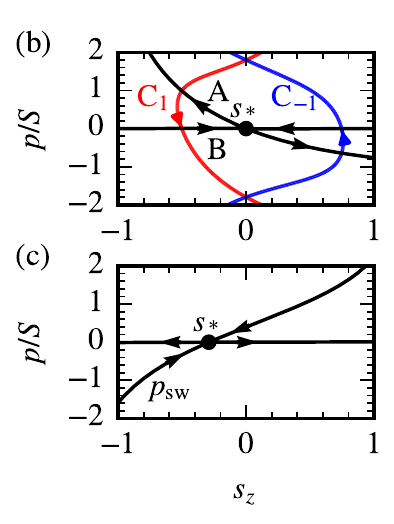}
  \caption{\label{fig:noise}
    (a)
    Noise in pumped charge current, for different tilt angles $\theta$
    and precession speeds $\Omega$, for $P_{F'}=P_{Fz}=0.9$.
    (b)
    Semiclassical trajectories for $V=-\Omega_{\mathrm{ext}}$,
    $P_{F'}=1$,
    $P_{Fz}=1/2$, $\chi=\xi=0$, $T=0$.
    Shown are the $H=0$ lines AB and the fixed point
    $s_*=\frac{1}{P_{Fz}P_{F'}}+\frac{2V}{P_{F'}\Omega}$ (black).
    Measurement trajectories $C_{\tilde{\chi}}$ for $\tilde{\chi}=1$ (red)
    and $\tilde{\chi}=-1$ (blue)
    are also shown.
    (c)
    Trajectories with anharmonicity, $\Omega=\Omega_{\mathrm{an}}s_z$,
    $V=1.5\Omega_{\mathrm{an}}$, $P_{Fz}=0.1$, $P_{F'}=1$, $T=0$.
  }
\end{figure}

The average dynamics follows the $\theta$ component of the
Landau-Lifshitz-Slonczewski equation, \cite{slonczewski1996-cde} here
obtained from stationarity of $S$ vs $\phi^q$,
\begin{align}
  \label{eq:spin-torque}
  {\cal S}
  \dot{\theta} = -\sin(\theta)\tau(\theta),
  \quad
  \tau(\theta)
  =
  \Omega \alpha(\theta) + I_{sz}
  \,,
\end{align}
where the spin current $I_{sz}=\frac{1}{4}G_0P_{Fz}V$ and damping
$\alpha(\theta)=\frac{1}{8}G_0[1-P_{Fz}P_{F'}\cos\theta]$
\footnote{
  For simplicity, we assume here that spin torque dominates magnetization damping.
  The presence of extra damping would lead to an additional term in $\alpha(\theta)$.
}
have been discussed in Ref.~\onlinecite{chudnovskiy2008-sts}.  The equation
describes motion of $\cos\theta$ in an effective potential
$-\int^{\cos\theta}\dd{(\cos\theta')}\tau(\theta')$ defined by $\Omega(\theta)$
and the spin torque, illustrated in
Fig.~\ref{fig:setup}b.  In certain parameter ranges, a fixed point
$\tau(\theta_*)=0$ appears --- it can be either
attractive or repulsive. This can correspond to a stable but
fluctuating precession around the angle $\theta_*$
(Fig.~\ref{fig:setup}b), induced by spin torque, or spin torque-induced
switching between two energy minima (Fig.~\ref{fig:setup}c). 

{\em Average current and noise.}  For fast
measurements, $t_0\ll{}1/\dot{\theta}$, we can assume $\theta$
remains fixed, and find the average currents,
\begin{align}
  \label{eq:avg-I}
  \overline{I} &=
  \frac{1}{2}G_0[1 + P_{F'}P_{Fz}\cos\theta]V + \frac{1}{4}G_0P_{Fz}\Omega \sin^2\theta
  \,,
  \\
  \overline{\dot{Q}_F}
  &=
  \frac{1}{2}\overline{I}V + \frac{1}{2}\tau(\theta)\Omega\sin^2\theta
  \,,
\end{align}
where the pumped charge current (second term in Eq.~\eqref{eq:avg-I}) is that found in
Ref.~\onlinecite{tserkovnyak2008-tbe}. The heat current is a sum of
the Joule heat and the magnetic energy lost due to the spin torque,
$\dot{E}_M=-\partial_t[\Omega_{\rm ext}S_z+\frac{1}{2}\Omega_{\rm an}S_z^2]$, dissipated
equally in $F$ and $F'$.  In contrast to the average values, the
energy shifts $\Omega_{\sigma\sigma'}$ remain in the noise of the
currents,
\begin{align}
  \label{eq:current-noise}
  S_I &=
  \sum_{\sigma\sigma'}\bar{G}_{\sigma\sigma'}
  \frac{1+\sigma\sigma'\cos\theta}{2}
  V_{\sigma\sigma'}\coth\frac{V_{\sigma\sigma'}}{2T}
  \,,
  \\
  S_{\dot{Q}_F}
  &=
  \sum_{\sigma\sigma'}\bar{G}_{\sigma\sigma'}
  \frac{1+\sigma\sigma'\cos\theta}{6}
  (\pi^2T^2 + V_{\sigma\sigma'}^2)V_{\sigma\sigma'}\coth\frac{V_{\sigma\sigma'}}{2T}
  \,,
\end{align}
where $V_{\sigma\sigma'}=V+\Omega_{\sigma\sigma'}$ and $T_F=T_{F'}$.
In the classical linear regime $V,\Omega<T$, the results reduce to a form dictated
by the fluctuation-dissipation theorem and Wiedemann-Franz law,
$S_I=2\bar{G}T$, $S_{\dot{Q}_F}=2\bar{G}L_0T^3$,
where $\bar{G}=\frac{\dd{\overline{I}}}{\dd{V}}$ is the electrical
dc conductance of the magnetic tunnel junction, \cite{huertas2002-asv}
and $L_0$ the Lorenz number.  The
presence of the angle-dependent frequencies is revealed in the quantum
noise regime $\Omega>T$.  The noise in the pumped current for $V=0$ is
plotted in Fig.~\ref{fig:noise} --- the location of the
quantum--classical crossover is pushed up to higher precession
frequencies as the tilt angle approaches $\theta=0$.

\begin{figure}
  \includegraphics{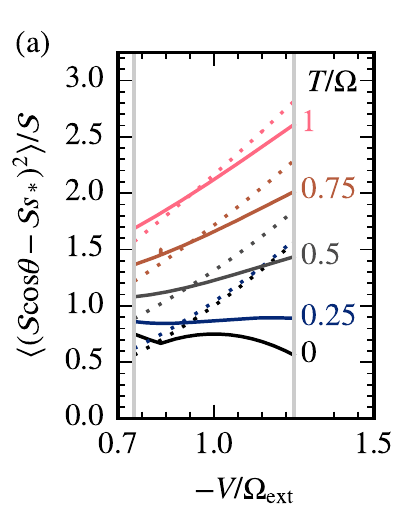}
  \includegraphics{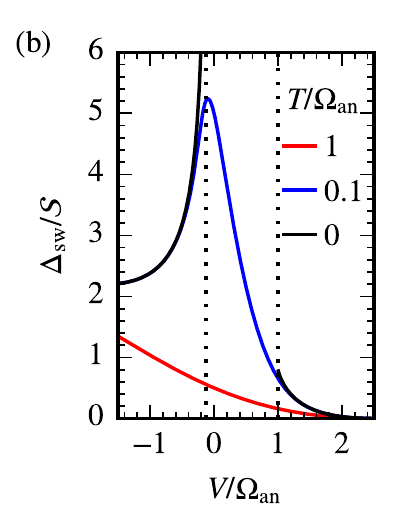}
  \caption{\label{fig:prob}
    (a)
    Normalized variance of the magnetization z-component in the steady state
    around the fixed point $s_*$, as a function of bias voltage and temperature for
    $P_{F'}=1/2$, $P_{Fz}=1/2$.
    Dotted lines indicate results where the energy shifts
    in the spin torque noise are neglected.
    (b)
    Switching exponent $\Delta_{\mathrm{sw}}$,
    for $P_{F'}=1$, $P_{Fz}=1/4$, and different temperatures and voltages.
    Dotted lines indicate the range $-\Omega_{\mathrm{an}}/8<V<\Omega_{\mathrm{an}}$
    where $\Delta_{sw}=\infty$ at $T=0$.
  }
\end{figure}

{\em Spin torque induced fluctuating precession.}  The above results
are conditional on a specific value of $\theta$.  For the full
probability distribution, the distribution $P(\theta)$ would need
to be known.

To find
$P(\theta)=\int\DD{\theta^{cl},\phi^q}e^{iS\rvert_{\chi=\xi=0}}\delta(\theta^{cl}(0)-\theta)$,
we assume ${\cal S}\gg1$ and take a semiclassical approximation.
Defining $s_z=\cos\theta$ and $p=2i{\cal S}\phi^q$, the action reads
$iS\rvert_{\chi=\xi=0}=\int\dd{t}[p\dot{s}_z - H(p,s_z)]$ where
$H=-iS_T$ is real for real $s_z$, $p$.  The problem can then be
analyzed as in Hamiltonian mechanics, $\dot{s}_z=\partial_pH$,
$\dot{p}=-\partial_{s_z}H$. \cite{kamenev2010-kta} In a time-sliced
discretization of the path integral, the $\delta$ restriction
specifying the exact measured value adds a boundary condition
$s_z(0)=s_{z0}$ that removes one of the integration variables and
saddle point equations. This allows for a discontinuity of $p$ at $t=0$,
cf. Refs.~\cite{pilgram2003-spi,heikkila2009-stf}.  The other boundary conditions are
$p(t\rightarrow\pm\infty)=0$, so that relevant paths have integration
constant $H=0$.

Consider now fluctuations close to an attractive fixed point
$\tau(\theta_*)=0$ (cf. Fig.~\ref{fig:setup}b). For dynamics driven by an
external field, it is located at
$s_z=s_*=\frac{1}{P_{F'}P_{Fz}}+\frac{2V}{P_{F'}\Omega_{\mathrm{ext}}}$,
and it is attractive if $\tau'(s_*)=-\Omega{}P_{Fz}P_{F'}<0$.  The
phase space picture is shown in Fig.~\ref{fig:noise}b.  Expanding
around $p=0$ in terms of the torque $\tau$ and torque noise correlator
$D$,
\begin{align}
  H &\simeq [1-s_z^2][{\cal S}^{-1}\tau(s_z)p - {\cal S}^{-2}D(s_z)p^2]
  \,,
  \\
  D(s_z)
  &=
  \frac{1}{8}
  \sum_{\sigma\sigma'\alpha=\pm}
  \frac{(1-\sigma\sigma's_z)^2}{1-s_z^2}
  \Gamma_{\sigma\sigma'\alpha}
  \,,
\end{align}
where
$\Gamma_{\sigma\sigma'\alpha}=\int_{-\infty}^\infty\dd{\epsilon}\Gamma_{\sigma\sigma'\alpha}(\epsilon)=\bar{G}_{\sigma\sigma'}\frac{1+\sigma\sigma's_z}{2}\alpha(V+\Omega_{\sigma\sigma'})/[1-e^{-\alpha(V+\Omega_{\sigma\sigma'})/T}]$
for $T_F=T_{F'}=T$. The fluctuation contribution comes from following
path A from $(s_*,0)$ to $(s_{z0},{\cal{}S}\tau(s_{z0})/D(s_{z0}))$:
\begin{align}
  \label{eq:sz-distribution}
  P(\cos\theta)
  \simeq
  N e^{{\cal S}\int_{s_*}^{\cos\theta}\dd{s_z}\frac{\tau(s_z)}{D(s_z)}}
  \simeq
  N e^{{\cal S}\frac{\tau'(s_*)}{D(s_*)}(\cos\theta-s_*)^2}
  \,,
\end{align}
where $N$ is a normalization constant. This agrees with
Ref.~\onlinecite{chudnovskiy2008-sts} in the semiclassical limit
${\cal S}\gg1$, except for the presence of the energy shifts
$\propto\Omega_{\sigma\sigma'}$  \cite{shnirman2015-gqn} in the spin torque noise correlator $D$,
which are relevant in
the quantum limit $\Omega\sim{}V\gg{}T$.  The variance is plotted in
Fig.~\ref{fig:prob}a.

{\em Long measurement times.}
For $t_0\gtrsim{}1/\dot{\theta}$, the slow
fluctuation of the magnetization contributes low-frequency noise to
observables. This contribution is not small in $1/{\cal S}$: the typical
excursion from the average position is small, $\delta{}s_z\propto{}{\cal
  S}^{-1/2}$, but it lasts for a long time $\tau_m\propto{}\mathcal{S}$,
generating low-frequency noise $S_{I}\sim{}(\frac{dI}{ds_z}\delta{}s_z)^2\tau_m$.
The situation is similar to noise induced in tunneling currents by
temperature fluctuations on small islands. \cite{laakso2010-fos}

\begin{figure}
  \includegraphics{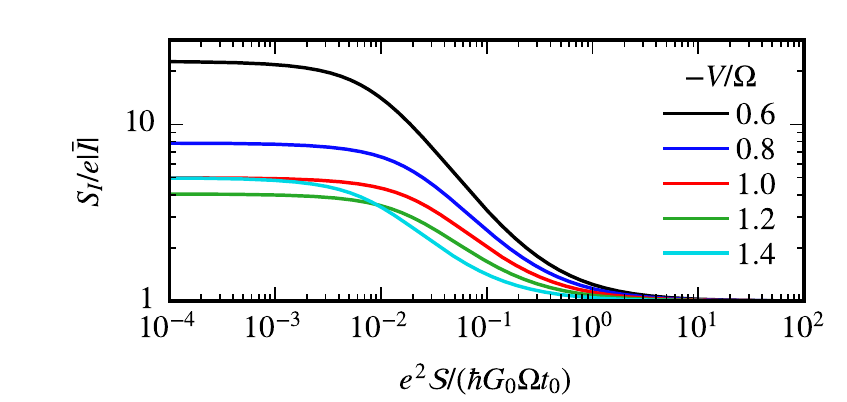}
  \caption{\label{fig:bandwidth}
    Current noise $S_I$ as a function of the measurement
    bandwidth $1/t_0$, for $P_{F'}=1$, $P_{Fz}=1/2$, $T=0$.
  }
\end{figure}

We now find the result within the semiclassical approximation.
The counting fields are switched on in the interval $0<t<t_0$,
e.g. $\chi(t)=\theta(t)\theta(t_0-t)i\tilde{\chi}$. They make the semiclassical
path to transition from branch A to B in the time interval $0<t<t_0$ following a trajectory
$\mathrm{C}_{\tilde{\chi},\tilde{\xi}}$ of constant
$H\rvert_{\tilde{\chi},\tilde {\xi}}$.
Two such trajectories are shown in Fig.~\ref{fig:noise}a.
For simplicity, we consider the limit $T\ll{}|\Omega|,|V|$ with full
polarization of the free magnet $P_{F'}=1$. Then, close to
$s_*$, %($|s_z-s_*|<|P_{Fz}|^{-1}-1$)
\begin{align}
  \label{eq:semiclH}
  H
  &\simeq
  e^{\alpha\tilde{\chi}}[1-s_*^2][
    \frac{\tau(s_z)}{{\cal S}}p
    - \frac{D(s_*)}{{\cal S}^2}p^2]
  - \Gamma(s_z)(e^{\alpha\tilde{\chi}}-1)
  \,,
\end{align}
where $\alpha=\sgn{V}$ and
$\Gamma(s_z)=\Gamma_{++\alpha}+\Gamma_{-+\alpha}$.
For quadratic $H$, the Hamiltonian equations can be solved exactly
(see \arxiv{Appendix}{Supplementary Material}).
From this approach, we find the current noise:
\begin{align}
  \label{eq:current-noise-2}
  S_I
  =
  \Gamma(s_*)
  +
  4\Gamma'(s_*)^2\sigma_s^2\tau_m
  \Bigl(1-\frac{1-e^{-t_0/\tau_m}}{t_0/\tau_m}\Bigr)
  \,,
\end{align}
where $\tau_m=-{\cal S}/[(1-s_*^2)\tau'(s_*)]$ is the slow time scale
associated with the spin transfer torque and
$\sigma_s^2=-D(s_*)/(2{\cal S}\tau'(s_*))$ the variance of the
magnetization z-component in Eq.~\eqref{eq:sz-distribution}.  The
first term $\Gamma(s_*)$  in Eq.~\eqref{eq:current-noise-2} is the
Poissonian shot noise~\eqref{eq:current-noise}, and the second term
originates from magnetization fluctuations.
The dependence on the measurement time is
shown in Fig.~\ref{fig:bandwidth}.  The current noise at frequencies
$\omega\sim\tau_m^{-1}\ll\Omega$ can be used to probe the
dynamics and distribution of the magnetization.

{\em Spin torque induced stochastic switching.}  Magnetic anisotropy
field $H_{\mathrm{an}}$ results to an effective magnetic potential
with two minima (see Fig.~\ref{fig:setup}c), and the spin torque
can induce switching between the two.
Here, we take $H_{\rm
  ext}=0$, and $\Omega=\gamma{}H_{\mathrm{an}}s_z\equiv\Omega_{\rm
  an}s_z$.  The corresponding semiclassical Hamiltonian picture is
shown in Fig.~\ref{fig:noise}c. An unstable fixed point
$s_*=\frac{1}{2P_{F'}P_{Fz}}[1-(1+8P_{F'}P_{Fz}^2V\Omega_{\mathrm{an}}^{-1})^{1/2}]$
separates the two stable fixed points $s_z=\pm1$.  The leading exponent of
the rate of switching from
$s_z=-1$ to $s_z=1$ is, \cite{utsumi2015-fts}
\begin{align}
  \Gamma_{\rm sw}
  &\propto
  e^{-\Delta_{\mathrm{sw}}}
  =
  e^{\int_{-1}^{s_*}\dd{s_z}p_{\mathrm{sw}}(s_z)}
  \,,
  &
  H(s_z, p_{\mathrm{sw}}(s_z)) &= 0
  \,,
\end{align}
where $p_{\mathrm{sw}}(s_z)$ is shown in Fig.~\ref{fig:noise}c. The
switching occurs deterministically ($\Delta_{\rm sw}\to0$) if
$P_{Fz}V>\frac{1+P_{F'}P_{Fz}}{2}\Omega_{\mathrm{an}}$ as $s_z=-1$
becomes unstable. At lower voltages, the switching is stochastic.
Numerically computed results are shown in Fig.~\ref{fig:prob}b.  At
zero temperature, the switching is blocked \cite{utsumi2015-fts} at
$-\frac{\Omega_{\mathrm{an}}}{8}<V<\Omega_{\mathrm{an}}$ for
$P_{F'}>P_{Fz}$ and
$-\Omega_{\mathrm{an}}<V<\frac{\Omega_{\mathrm{an}}}{8}$ otherwise.
This occurs because the transition rates
$\Gamma_{\sigma\sigma'\alpha}$ vanish for
$\alpha(V+\Omega_{\sigma\sigma'})\le0$, and because the back-action
$\propto\Omega_{\sigma\sigma'}$ vanishes for $\sigma=-\sigma'$,
$s_z\to-1$.
\footnote{
  The blocking is due to non-Gaussianity of the spin torque.
  The result neglects the exact quantization of spin and ignores
  e.g. quantum tunneling.
}
The latter constraint is due to the additional angle
dependence in the spin torque, which traces back to the geometric
phases \cite{shnirman2015-gqn} in the spin dynamics.

{\em Discussion.}
In conclusion, we have derived a Keldysh action~\eqref{eq:genfunc},
describing the stochastic charge and energy currents affected by a
precessing magnetization.  We obtain a fluctuation relation for the
transferred charge, energy, and magnetization.  The noise in the
current at low temperatures displays features related to geometric
phases, and its low frequency component reflects the magnetization
fluctuations. Information about the spin torque noise is also
contained in the switching probability of anisotropic magnets.  Our
predictions are readily accessible in experiments probing spin pumping
at low temperatures $T < \hbar \Omega/k_B$. Precession frequencies in
$\unit[10]{GHz}$ range have been achieved, \cite{kiselev2003-mon,rippard2004-dci}
which translates to $T \lesssim \unit[1]{K}$.

We thank B. Nikolic and S. van Dijken for discussions.
This work was supported by the MIUR-FIRB2013 - Project Coca (Grant
No. RBFR1379UX), the Academy of Finland Centre of Excellence program
(Project No. 284594) and the European Research Council (Grant
No. 240362-Heattronics).

\clearpage

\onecolumngrid
\appendix

\section{\arxiv{Appendix: Details}{Supplementary material: Details} of derivation of the generating function}

We consider a tunneling Hamiltonian model for the ferromagnet/nanomagnet junction,
\begin{align}
  \label{eq:hamiltonian}
  H &= \sum_{\vec{k}\vec{k}'\sigma\sigma'}[
    c^\dagger_{\sigma\vec{k}}({\cal H}_{F'})_{\sigma\vec{k},\sigma'\vec{k}'} c_{\sigma'\vec{k}'}
    +
    d^\dagger_{\sigma\vec{k}} ({\cal H}_F)_{\sigma\vec{k},\sigma'\vec{k}'} d_{\sigma'\vec{k}'}
  ]
  +
  H_T
  +
  H_{\rm ext}
  \,,
  \\
  ({\cal H}_{F'})_{\sigma\vec{k},\sigma'\vec{k}'}
  &=
  \delta_{\vec{k},\vec{k}'}[
    1\epsilon_{F',\vec{k}} + g\vec{S}\cdot\vec{\sigma}
  ]_{\sigma\sigma'}
  \,,
  \\
  ({\cal H}_{F})_{\sigma\vec{k},\sigma'\vec{k}'}
  &=
  \delta_{\vec{k},\vec{k}'}\epsilon_{F,\sigma\sigma'\vec{k}}
  \,,
  \\
  H_T
  &=
  \sum_{\sigma\sigma'\vec{k}\vec{k'}} W_{\sigma\vec{k},\sigma'\vec{k}'}
  c_{\sigma\vec{k}}^\dagger d_{\sigma'\vec{k}'}
  +
  \mathit{h.c.}
  \,,
  \\
  H_{\rm ext}
  &=
  J \vec{H}_{\rm ext}\cdot\vec{S}
  \,.
\end{align}
Above, $c$ ($d$) are conduction electrons in the free (fixed) magnet,
$\vec{S}$ is the magnetization in the free (single-domain) magnet,
$\vec{H}_{\rm ext}$ an externally applied field, and $g$ and $J$
coupling constants. Moreover, $\epsilon_{F'/F}$ describe the
noninteracting energy dispersions.

The Keldysh action corresponding to $H$ is,
\begin{align}
  S &=
  S_0[\vec{S}] + \int_{-\infty}^\infty\dd{t}\bigl(\bar{c}^T [
  i\partial_t - {\cal H}_{F'}] c + \bar{d}^T [i\partial_t - {\cal H}_F] d
  + \bar{c}^TW d + \bar{d}^T W^\dagger c\bigr)
  \,,
\end{align}
where $S_0$ is the standard spin action \cite{abanov2002-bpf}.
We also include source terms in the generating function $Z$ \cite{kindermann2004-sht},
\begin{align}
  Z[\chi,\xi,\zeta]
  &=
  \int\DD{\vec{S},\bar{c},c,\bar{d},d}e^{iS+iS_c+iS_{c2}}
  \,,
  \\
  S_c[\chi,\xi] &=
  \frac{1}{2}
  \int_{-\infty}^\infty\dd{t}\bar{d}^T\check{\gamma}_x[
    \dot{\chi}
    +
    ({\cal H}_F - \mu) \dot{\xi}
  ]d
  \,,
  \\
  \label{eq:spincount}
  S_{c2}[\zeta] &= \int_{-\infty}^\infty\dd{t} \zeta \partial_t S_z^{cl}
\end{align}
so that derivatives vs. $\chi$ and $\xi$ produce cumulants of the
charge and energy transfer, and $\zeta$ characterizes $S_z$.  The
terms added in $S_c$ can be eliminated with a change of variables
$d(t)\mapsto{}e^{-i\gamma_x[({\cal H}_F-\mu)\xi(t) + \chi(t)]/2}d(t)$
and conversely for $\bar{d}$, provided $\chi(\pm\infty)=\xi(\pm\infty)=0$.
This results to addition of phase
factors in $W$, via $W\mapsto{}We^{-i\gamma_x[({\cal
      H}_F-\mu)\xi(t)+\chi(t)]/2}$. Here and below, we use
Larkin-Ovchinnikov rotated Keldysh basis \cite{kamenev2010-kta}:
fermion fields have Keldysh structure $d=(d_1,d_2)$, where
$d_{1/2}=(d_+\pm{}d_-)/\sqrt{2}$,
$\bar{d}_{1/2}=(\bar{d}_+\mp\bar{d}_-)/\sqrt{2}$ are related to the
fields $d_\pm$ on the Keldysh branches. Real fields are split
similarly, as $X^{cl/q}=(X_+\pm{}X_-)/2$. $\check{\gamma}_x$ is a
Pauli matrix in the Keldysh space.

We integrate out the conduction electrons, and expand in small $W$,
$\dot{\vec{S}}$ \cite{shnirman2015-gqn,chudnovskiy2008-sts}. The
resulting tunneling action can be obtained via the same route as in
Ref.~\onlinecite{shnirman2015-gqn},
\begin{align}
  S_T[\chi,\xi,\vec{S}]
  =
  i
  \int\dd{t}\dd{t'}
  \Tr[
  R(t)G_{F'}(t,t')R(t')^\dagger
  W e^{-i\gamma_x[{\cal H}_F\xi(t')+\chi(t')]/2} G_F(t',t) e^{i\gamma_x[{\cal H}_F\xi(t)+\chi(t)]/2}  W^\dagger
  ]
  \,.
\end{align}
Here, $G_{F'}^{R/A} = [\epsilon\pm{}i0^+ - \xi_{F'} - g
  |\vec{S}|\sigma_z]^{-1}$,
$G_F^{R/A}=[\epsilon\pm{}i0^+-\xi_N]^{-1}$, and the unitary matrices
$R$ are defined by $R\sigma_zR^\dagger=\vec{S}\cdot\vec{\sigma}$ and
the gauge degree of freedom $\psi$ in $R=R'e^{i\psi\sigma_z}$ is fixed
\cite{shnirman2015-gqn} so that $(R^\dagger\partial_t R)^{cl}_z=0$ and
$(R^\dagger\partial_t R)^{q}_z$ is proportional to time derivatives of
classical field components.  The conduction electrons $c$ may also
contribute other terms than $S_T$, for example change (or generate) the
total spin in $S_0$ \cite{abanov2002-bpf}. However, here we are
mainly interested in spin torque and pumping, and therefore concetrate
on dynamics implied by $S_T$ and assume any other effects are absorbed
to changes in parameters or phenomenological damping terms.

Let us now consider the long-time limit correlation
functions of the form
$\Tr[e^{i\xi_0O(t_0)/2}e^{-i\xi_0O(0)/2}\rho{}e^{-i\xi_0O(0)/2}e^{i\xi_0O(t_0)/2}]$, 
$t_0\to\infty$, which characterize a two-measurement
protocol \cite{esposito2009-nff,campisi2011-cqf}.
They correspond to choices
$\xi(t)=\theta(t)\theta(t_0-t)\xi_0$, and
$\chi(t)=\theta(t)\theta(t_0-t)\chi_0$.
We can write (see below)
\begin{align}
  \label{eq:energy-counting}
  &e^{-i\gamma_x{\cal H}_N\xi(t)}
  G_F(t,t')
  e^{i\gamma_x{\cal H}_N\xi(t')}
  %\\\notag
  %&
  =
  \int_{-\infty}^\infty\frac{\dd{\epsilon}}{2\pi}
  e^{-i\epsilon(t-t')}
  e^{-i\epsilon\gamma_x\xi(t)}G_F(\epsilon)e^{i\epsilon\gamma_x\xi(t')}
  +
  a(t,t')
  \,,
\end{align}
where the correction term $a(t,t')$ is zero for
$|t-t'|\ge{}|\xi(t)-\xi(t')|$.  As discussed below in more detail, it can be neglected in the
long-time limit \cite{kindermann2004-sht}. The tunneling action then
reads
\begin{align}
  \label{eq:longtimelimit}
  S_T
  &\simeq
  i\int_{-\infty}^\infty\dd{t}\int_{-\infty}^\infty\frac{\dd{\epsilon}}{2\pi}
  \Tr[
    \overline{
      P(t,\epsilon)G_{F'}(t,t')P(t',\epsilon)^\dagger
    }
    W G_F(\epsilon)W^\dagger
  ]
  \,,
\end{align}
where $P(t,\epsilon) =
e^{i(\epsilon-V)t}e^{-i\gamma_x\chi(t)/2}e^{-i\gamma_x\epsilon\xi(t)/2}R(t)$, and
$\overline{X(t,t')}=\int_{-\infty}^\infty\dd{t'}X(t+t'/2,t-t'/2)$.

In the case considered in the main text, $\check{\phi}=\Omega t +
\check{\tilde{\phi}}$, and dynamics of $\check{\theta}$ and
$\check{\tilde{\phi}}$ arises from the spin transfer torque. We have
\begin{align}
  R(t)
  &=
  e^{-i[\Omega{}t + \psi_0(t)]\sigma_z/2}
  e^{-i\sigma_z[\check{\tilde{\phi}} + \check{\tilde{\psi}}]/2}
  \cos\frac{\check{\theta}}{2}
  +
  e^{-i[\Omega{}t - \psi_0(t)]\sigma_z/2}
  e^{-i\sigma_z[\check{\tilde{\phi}} - \check{\tilde{\psi}}]/2}
  (-i\sigma_y)\sin\frac{\check{\theta}}{2}
  \,,
\end{align}
where $\psi_0(t)=-\int^t\dd{t'}\Omega\cos\theta^{cl}$,
$\psi_0(t+q)\simeq{}\psi_0(t) - \Omega{}q\cos\theta^{cl}(t)$.
Keeping only the
non-oscillating parts of Eq.~\eqref{eq:longtimelimit} and taking
the leading term of the gradient expansion vs. $\theta$, $\tilde{\phi}$, we
can write the time average:
\begin{align}
  &\overline{
    e^{i\epsilon(t-t')}
    P(t,\epsilon)G_{F'}(t,t')P(t',\epsilon)^\dagger
  }
  \simeq
  \int_{-\infty}^\infty\dd{\epsilon'}
  \sum_{s=\uparrow/\downarrow}
  \sigma_{s}
  \tr
  \Bigl(
  \\\notag&\qquad
  \delta(\epsilon-\epsilon'-V-s\Omega_-(t))
  e^{i\eta_{-,s}(\epsilon,t)\check{\gamma}_x/2}
  \cos\frac{\check{\theta}(t)}{2}
  \check{G}_{1,s}(\epsilon')
  \cos\frac{\check{\theta}(t)}{2}
  e^{-i\eta_{-,s}(\epsilon,t)\check{\gamma}_x/2}
  \\\notag&\qquad
  +
  \delta(\epsilon-\epsilon'-V-s\Omega_+(t))
  e^{i\eta_{+,s}(\epsilon,t)\check{\gamma}_x/2}
  \sin\frac{\check{\theta}(t)}{2}
  \check{G}_{1,-s}(\epsilon')
  \sin\frac{\check{\theta}(t)}{2}
  e^{-i\eta_{+,s}(\epsilon,t)\check{\gamma}_x/2}
  \Bigr)
  \,,
\end{align}
where $\eta_{\pm,s}(\epsilon,t)
=\chi(t) + [\epsilon - \mu]\xi(t) - s[\tilde{\phi}^q(t) \mp \tilde{\psi}^q(t)]=\chi(t) + [\epsilon - \mu]\xi(t) - 2s\frac{\Omega_\pm(t)}{\Omega}\tilde{\phi}^q(t)$ and $\Omega_{\pm}=\Omega[1\pm\cos\theta^{cl}]/2$.
To this order, $S_T$
is independent of $\tilde{\phi}^{cl}$. Provided no source fields measuring
the statistics of $\tilde{\phi}^{cl}$ are added, the only part dependent on it
is $S_0=\ldots+{\cal
  S}\int\dd{t}\tilde{\phi}^{cl}\frac{1}{2}\partial_t[\cos(\theta^{cl}+\theta^q)-\cos(\theta^{cl}-\theta^q)]$, which implies a constraint $\cos(\theta^{cl}+\theta^q)-\cos(\theta^{cl}-\theta^q)=\text{const.}$ and we set $\theta^q=0$.  The slow part of the
dynamics of the polar $\phi$ angle decouples from the rest of the
problem.

The result Eq.~(3) in the main text now
follows, noting
$\int\dd{t}\dd{t'}(\ldots)G^R_{F'}(t,t')G^R_F(t',t)=\int\dd{t}\dd{t'}(\ldots)G^A_{F'}(t,t')G^A_F(t',t)=0$
and neglecting terms unimportant in the long-time limit introduced by Eq.~\eqref{eq:energy-counting}.

\section{Energy counting}

In the eigenbasis of the single-particle operator ${\cal H}_N$, we can
write
\begin{align}
  \check{G}_F(\epsilon)
  &=
+  \sum_j\check{G}_{F,j}(\epsilon)\ket{j}\bra{j}
  \,,
  \\
  \check{G}_{F,j}(\epsilon)
  &=
  \frac{\cal P}{\epsilon - \epsilon_j}
  +
  i\pi\delta(\epsilon-\epsilon_j)
  \begin{pmatrix}
    1
    &
    2\tanh\frac{\epsilon-\mu}{2T}
    \\
    0
    &
    -1
  \end{pmatrix}
  \,,
\end{align}
where ${\cal P}$ is the Cauchy principal value. Straightforward
calculation now yields
\begin{equation}
\begin{split}
  &e^{i{\cal H}_F\check{\gamma}_x\xi(t)}\check{G}_F(t,t')e^{-i{\cal H}_F\check{\gamma}_x\xi(t')}
  \\
  &=
  \int\frac{\dd{\epsilon}}{2\pi}
  e^{-i\epsilon[t+\check{\gamma}_x\xi(t)]}
  \check{G}_F(\epsilon)
  e^{i\epsilon[t' + \check{\gamma}_x\xi(t')]}
  +
  \int\frac{\dd{\epsilon}}{2\pi}
  \sum_j
  e^{-i\epsilon(t-t')}
  e^{i\epsilon_j\gamma_x[\xi(t) - \xi(t')]}
  [
    1
    -
    e^{i(\epsilon-\epsilon_j)\gamma_x[\xi(t) - \xi(t')]}
  ]
  \frac{\cal P}{\epsilon - \epsilon_j}
  \ket{j}\bra{j}
  \\
  &=
  \int\frac{\dd{\epsilon}}{2\pi}
  e^{-i\epsilon[t+\check{\gamma}_x\xi(t)]}
  \check{G}_F(\epsilon)
  e^{i\epsilon[t' + \check{\gamma}_x\xi(t')]}
  +
  \sum_j
  e^{-i\epsilon_j(t-t' + \gamma_x\xi(t)-\gamma_x\xi(t'))}
  \ket{j}\bra{j}
  \frac{-i}{4}
  [a_1 + a_2\gamma_x]
  \\
  &
  a_1(t,t')
  =
  2\sgn(t-t')
  -\sgn(t-t' + \xi(t) - \xi(t'))
  -\sgn(t-t' - \xi(t) + \xi(t'))
  \,,
  \\
  &a_2(t,t')
  =
  \sgn(t-t' + \xi(t) - \xi(t'))
  -
  \sgn(t-t' - \xi(t) + \xi(t'))
  \,.
\end{split}
\end{equation}
The correction term is zero for $|t-t'|\ge|\xi(t)-\xi(t')|$, and
consequently gives negligible contribution in the long-time limit
where $\xi(t)=\xi(t')$ except near the ends of the measurement
interval.  While neglecting it is not necessary in principle, this
simplifies the approach. Appearance of such ``time-energy uncertainty''
was also noted in Ref.~\cite{kindermann2004-sht}.

\section{Fluctuation theorem}
Let us consider a system where the nanomagnet is coupled to a
ferromagnetic electrode via a tunnel barrier with spin-flip
conductance $G_T$ and polarization $P$, and via an ohmic contact to a
normal metal. We disregard the magnetization damping caused by the
normal metal, and assume that the voltage completely drops across the
tunnel junction. Below, we denote the temperature of the normal metal
and the nanomagnet by $T_{F'}$, and that of the ferromagnetic
electrode by $T_F$. We disregard charge and energy pile-up
effects, limiting ourselves to time scales long compared to the charge
and energy relaxation time of the system. In this case we
can specify the probability distribution of charge $\delta n=\int_t^{t+t_0} dt'
I(t'))$ and energy $\delta E=\int_t^{t+t_0} dt' \dot Q(t')$ tunneling
through the tunnel contact, and a change in the $z$
component of magnetization $\delta M_z={\mathcal S}\gamma \delta s_z/\nu$,
$\delta s_z=\delta [\cos(\theta^{\rm cl})]$ in a
time $t_0$. It reads 
\begin{equation}
\label{eq:prob}
\begin{split}
&P_{\rm t_0}(\delta n,\delta E,\delta s_z) \\&= \int
\frac{d\zeta}{2\pi} \frac{d\chi_{0}}{2\pi}
 \frac{d\xi_{0}}{2\pi}  e^{-i\chi_{0} \delta n - i \xi_{0}
  \delta E} {\cal D} \phi^q {\cal D} s_z e^{\left\{i \int S\zeta(t) \left[
\partial_{t} s_z(t)/{\cal V} -\frac{\delta s_z}{t_0}
\right]dt \right\}} e^{i
[S_0(\phi^q)+S_{T}(\chi,\xi,\phi^q;V,T_F,T_{F'})]}\\&= \int
\frac{d\zeta}{2\pi} \frac{d\chi_{0}}{2\pi}
\frac{d\xi_{0}}{2\pi}  {\cal D} \phi^q {\cal D} s_z e^{-i\chi_{0} \delta n - i \xi_{0}
  \delta E-i \zeta \delta s_z}
e^{i\{S_0[\phi^q(t)+\zeta(t)/2]+S_{T}(\chi,\xi,\phi^q;V,T_F,T_{F'})\}},
\end{split}
\end{equation}
where $\chi(t)=\chi_0\theta(t) \theta(t_0-t)$, $\xi(t)=\xi_0\theta(t)
\theta(t_0-t)$ and $\zeta(t)=\zeta_0\theta(t)\theta(t_0-t)$ specify
the two-measurement protocol. Since $S_T$ is independent of the slow component of the
$\phi$-coordinate, this component does not affect the statistics of
the other parameters and can be integrated out. In the above equation,
we assume that the measurement time $t_0$ is long compared to charge
relaxation times of the island, but it can be short compared to the
time scale of magnetization relaxation. The presence of the spin action
$S_0$ ensures the conservation of the total angular momentum, analogous to
the other conservation laws explained in \cite{pilgram04}. 

We can use the Fermi function identities $f(2\mu-\epsilon)=f(\epsilon)e^{\beta(\epsilon-\mu)}=1-f(\epsilon)$ to
get 
\begin{equation}
\Lambda_+(\epsilon,V)=e^{(\epsilon-\mu)/T_{F'}}
e^{-(\epsilon-\mu+V)/T_F}\Lambda_-(\epsilon,V), \quad
\Lambda_+(2\mu-\epsilon,-V)=\Lambda_-(\epsilon,V).
\end{equation}
Applying these to the action $S_T$ yields the symmetries
\begin{subequations}
\begin{align}
S_T(\chi,\xi,\phi^q;V,\Omega,T_F,T_{F'})
  &=S_T(-\chi,\xi,-\phi^q;-V,-\Omega,T_F,T_{F'})\label{eq:rela}\\
S_T(\chi,\xi,\phi^q;V,\Omega,T_F,T_{F'}) 
&=S_T(-\chi-\xi V,-\xi,-\phi^q-\xi \Omega/2;-V,-\Omega,T_{F'},T_F) \label{eq:relb}\\
S_T(\chi,\xi,\phi^q;V,\Omega,T_F,T_{F'})
  &=S_T(-\chi+iV/T_F,-\xi+i(T_F^{-1}-T_{F'}^{-1}),-\phi^q+i
    \Omega/(2T_F);V,\Omega,T_F,T_{F'}) \label{eq:relc}
\end{align}
\end{subequations}
In addition, one more relation can be obtained by reversing the
magnetizations of both systems, or changing the signs of the
polarizations $P_{Fz}$ and $P_{F'}$. This sign change can be balanced
by replacing $\Omega \mapsto -\Omega$ and $\phi^q \mapsto
-\phi^q$. Denoting the magnetization reversal and reversal of the sign
of precession with a prime hence
yields the symmetry
\begin{equation}
S_T(\chi,\xi,\phi^q;V,\Omega,T_F,T_{F'})=S_T'(\chi,\xi,-\phi^q;V,\Omega,T_F,T_{F'}). \label{eq:reld}
\end{equation}
These relations together with the definition \eqref{eq:prob} allow us
to find various symmetries of the probability distribution. For
example, combining \eqref{eq:rela} with \eqref{eq:relb} yield
\begin{equation}
\label{eq:prob1}
\begin{split}
P_{\rm t_0}(\delta n,\delta E,\delta s_z;T_F,T_{F'}) % \\&= \int
=P_{\rm t_0}(\delta n,-\delta E+V \delta n-\delta s_z \Omega,\delta s_z,T_{F'},T_F).
\end{split}
\end{equation}
This relates the probabilities of charge and energy transfer and
change of magnetization upon the interchange of the two
temperatures. The detailed form is a result of the particle-hole
symmetry of our model (no thermoelectric effects are included). For
$T_F=T_{F'}$ this implies the first law of thermodynamics for the
processes. Namely, it implies
\begin{equation}
2 \langle \delta E \rangle - V \langle \delta n \rangle + \Omega
\langle \delta s_z \rangle = 0,
\end{equation}
i.e., for an arbitrary nonequilibrium state, the expectation value of
the internal energy increase in the two terminals  (when $T_F=T_{F'}$,
$\langle \delta E_F\rangle = \langle \delta E_{F'}\rangle \equiv
\langle \delta E\rangle$) equals the sum of
the dissipated Joule heat and the work done by the change in the
magnetization direction.
This result is also reflected in the average heat current in
Eq.~(9) of the main text.

On the other hand, combining \eqref{eq:relc} with \eqref{eq:reld} yields
\begin{equation}
\label{eq:prob2}
\begin{split}
  P_{\rm t_0}(\delta n,\delta E,\delta s_z;T_F,T_{F'})
=e^{V \delta n/T_F}
e^{\delta E (T_F^{-1}-T_{F'}^{-1})} e^{- \delta s_z
  \Omega/T_F} P_{\rm t_0}'(-\delta n,-\delta E,\delta
s_z;T_F,T_F').
\end{split}
\end{equation}
This is the spintronic fluctuation relation for the setup, and it in
particular cases yields those presented in
\cite{seifert2012-stf,crooks1999-epf,crooks2000-pea,tobiska05,utsumi2015-fts}.
Note that in contrast to \cite{utsumi2015-fts}, this presents the
probability statistics of change of magnetization, rather than that of
the spin current, as the latter as such is difficult to measure directly. 

One direct consequence of fluctuation theorems is the Onsager symmetry
of linear response coefficients characterizing nonequilibrium
observables \cite{andrieux04,esposito2009-nff}. In particular, we can define the electrical and energy currents and
spin transfer torque via
\begin{subequations}
\begin{align}
I_c &= \frac{\partial S_T}{\partial \chi}|_{\chi=\xi=\phi_q=0}\\
\dot Q &= \frac{\partial S_T}{\partial\xi}|_{\chi=\xi=\phi_q=0}\\
\sin^2(\theta) \tau &= \frac{\partial S_T}{\partial \zeta}|_{\chi=\xi=\phi_q=0},
\end{align}
\end{subequations}
or denoting $\lambda_{1,2,3} \in \{\chi,\xi,\phi_q\}$ and generalized
current as
\begin{equation}
J_i \equiv \frac{\partial S_T}{\partial \lambda_i}|_{\lambda=0}.
\end{equation}

Let us consider $f_{1,2,3} \in \{V,T_F(T_F^{-1}-T_{F'}^{-1}),\Omega/2\}$ as
generalized forces. The linear response coefficients are thus defined via
\begin{equation}
L_{ij} \equiv \frac{\partial J_i}{\partial f_j}|_{\vec f=0} =
\frac{\partial^2 S_T}{\partial\lambda_i \partial f_j}|_{\vec
  \lambda=\vec f=\vec 0}.
\end{equation}
In terms of the generalized forces and currents, Eq.~\eqref{eq:relc} can be expressed as
\begin{equation}
S_T(\vec \lambda;\vec f)
  =S_{T}(-\vec \lambda+i \vec f/T_F;\vec f).
\end{equation}
Differentiating this with respect to $f_j$ and $\lambda_i$ and
setting $\vec \lambda=\vec f=0$ then gives (now
$T_F=T_{F'} \equiv T$)
\begin{equation}
2T L_{ij} \equiv 2T \frac{\partial S_T}{\partial \lambda_i \partial f_j}(0;0) = -i \frac{\partial^2 S_T}{\partial \lambda_i \partial
  \lambda_j}(0;0).\label{eq:Lijsymm}
\end{equation}
This hence yields the Onsager reciprocity relations, accoring to which
$L_{ij}=L_{ji}$ is a symmetric matrix. Note that
Eq.~\eqref{eq:Lijsymm} is the (zero-frequency) fluctuation-dissipation theorem:
The left hand side of this equation yields the zero-frequency
autocorrelation function $\langle \delta J_i \delta J_j\rangle$.

\section{Semiclassical equations}
In the main text we analyze the slow stochastic dynamics of the magnetization
under the fluctuating spin torque within the semiclassical
approximation. As a result, the $z$ component of the magnetization of
the free magnet, $s_z=\cos(\theta)$, and a term proportional to the
counting field $\phi^q$, $p \equiv 2i{\cal S} \phi^q$ become conjugate
variables, whose dynamics follows the effective Hamiltonian
$H=-iS_T$. Let us consider such Hamiltonian mechanics generated by a
quadratic Hamiltonian, such as that in Eq.~(15) in the main text, $H=-A qp - B p^2 + C + D q + E
q^2$ with initial and final conditions $B p(0) + A q(0) = 0$ and
$p(t_0)=0$.  The integration constant is $H_0=C + D q(0) + E q(0)^2$,
which also implies $q(t_0)=q(0)$.

Define $r = p + \alpha q$ where $\alpha=\frac{A+\beta}{2B}$ and
$\beta=\sqrt{A^2+4BE}$.  Then,
\begin{align}
  H &= \beta q r - B r^2 + C + D q
  \,,
  &
  \dot{r} &= -\beta r - D
  \,.
\end{align}
Taking the initial and final conditions into account,
we find the starting point of the trajectory
\begin{align}
  q(0) = \frac{-D}{\alpha \beta + \frac{A/B}{e^{\beta t_0} - 1}}
  \,,
\end{align}
so that
\begin{align}
  r(t)
  =
  e^{-\beta t}\frac{\beta-A}{2B}q(0) - \frac{D}{\beta}[1 - e^{-\beta t}]
  \,.
\end{align}
The action of the trajectory now reads
\begin{align}
  iS_C
  =
  \int_0^{t_0}\dd{t}[p\dot{q} - H]
  =
  -t_0H_0 + \int_{0}^{t_0}\dd{t}p\dot{q}
  \,.
\end{align}
The second term is ($A,B>0$)
\begin{align}
  \int_{0}^{t_0}\dd{t}p\dot{q}
  &=
  -p(0)q(0) - \int_{0}^{t_0}\dd{t}q[\dot{r} - \alpha \dot{q}]
  =
  \frac{Aq(0)^2}{B}
  +
  \int_{r(0)}^{r(t_0)}\dd{r}\frac{H_0 - C + B r^2}{D + \beta r}
  \\
  &=
  \frac{Aq(0)^2}{B}
  -
  \frac{B}{2\beta^2}\Bigl[(\beta r(t_0) - 2D)r(t_0)
  -(\beta r(0) - 2D)r(0)\Bigr]
  +
  t_0[H_0 - C + \frac{BD^2}{\beta^2}]
\end{align}
For $\beta{}t_0\gg1$, the last term is $\propto t_0 e^{-\beta t_0}$,
and the others approach a constant as $t_0\to\infty$.
Therefore, for $t_0\to\infty$ we have $iS_C\simeq{}-t_0H_0$.

Let us furthermore expand the action in a counting field $\lambda$ to
the second (Gaussian) order. Given expansions $A=A_0+A_1\lambda +
A_2\lambda^2 + \ldots$, of $A$, 
$B$, $C$, $D$, $E$, with $C_0=D_0=E_0=0$, we find the
leading terms
\begin{align}
  iS_C
  &\simeq
  - C_1 \lambda t_0
  - C_2 \lambda^2 t_0
  +
  \lambda^2
  \frac{B_0D_1^2}{A_0^2}\Bigl[t_0 - \frac{1 - e^{-2A_0t_0}}{2A_0}\Bigr]
  \,,
  \\
  iS_A
  &=
  \int_{-\infty}^0\dd{t}p\dot{q}\rvert_{C=D=E=0}
  =
  -\frac{Aq(0)^2}{2B}
  \simeq
  -
  \lambda^2
  \frac{B_0D_1^2}{2A_0^3}(1 - e^{-A_0 t_0})^2
  \,.
\end{align}
Adding $iS=iS_A+iS_C$, taking $\lambda=\alpha\tilde{\chi}$, and substituting
in the values corresponding to Eq.~(15) of the main text,
$A_0=-[1-s_*^2]\tau'(s_*)/{\cal S}$,
$B_0=[1-s_*^2]D(s_*)/{\cal S}^2$,
$C_1=-\alpha\tilde{\chi}\Gamma(s_*)$,
$C_2=-\frac{1}{2}\tilde{\chi}^2\Gamma(s_*)$, $D_1=-\Gamma'(s_*)\alpha\tilde{\chi}$,
we arrive at Eq.~(16) in the main text. 

\end{document}